\documentclass[10pt,letterpaper]{article}
\usepackage{opex3}
\usepackage{caption}
\usepackage{subcaption}
\usepackage{graphicx}
\usepackage{csquotes}
\usepackage{amsmath}
\usepackage{amssymb}

\newcommand{\ket}[1]{|#1 \rangle}
\newcommand{\bra}[1]{\langle #1|}

\begin{document}

%%%%%%%%%%%%%%%%%% title page information %%%%%%%%%%%%%%%%%%
\title{Long-range coupling of silicon photonic waveguides using lateral leakage and adiabatic passage}

\author{A. P. Hope$^{1}$, T. G. Nguyen$^{1}$, A. D. Greentree$^{2}$, A. Mitchell$^{1}$}
\address{$^{1}$ARC Centre of Excellence for Ultrahigh bandwidth Devices for Optical systems (CUDOS)\\
 and School of Electrical and Computer Engineering, RMIT University, Melbourne, Australia\\
$^{2}$Applied Physics, School of Applied Sciences, RMIT University, Melbourne, Australia\\
}

\email{*anthony.hope@rmit.edu.au} %% email address is required

%%%%%%%%%%%%%%%%%%% abstract and OCIS codes %%%%%%%%%%%%%%%%
\begin{abstract}
We present a new approach to long range coupling based on a combination of adiabatic passage and lateral leakage in thin shallow ridge waveguides on a silicon photonic platform.  The approach enables transport of  light between two isolated waveguides through a mode of the silicon slab that acts as an optical bus. Due to the nature of the adiabatic protocol, the bus mode has minimal population and the transport is highly robust. 
We prove the concept and examine the robustness of this approach using rigorous modelling.
We further demonstrate the utility of the approach by coupling power between two waveguides whilst bypassing an intermediate waveguide. This concept could form the basis of a new interconnect technology for silicon integrated photonic chips.
\end{abstract}

 % REPLACE WITH CORRECT OCIS CODES FOR YOUR ARTICLE

%%%%%%%%%%%%%%%%%%%%%%% References %%%%%%%%%%%%%%%%%%%%%%%%%
\bibliographystyle{osajnl}

%%%%%%%%%%%%%%%%%%%%%%%%%%  body  %%%%%%%%%%%%%%%%%%%%%%%%%%
%\newpage
\section{Introduction}
Mass manufacture of monolithic systems of extraordinary complexity, compactness and precision using CMOS processing has underpinned the information revolution.  However, interconnections between complex functional blocks remains a critical challenge, often requiring numerous interconnect layers above the functional plane. The CMOS process has recently been adapted to photonic integrated circuits \cite{Bogaerts:05} with applications emerging in high-speed communications \cite{Ding:12}, photonic signal processing \cite{Koos:09} and quantum optics \cite{Politi:08}. Silicon photonic systems are gaining momentum but device complexity will again be limited by interconnect technology. Out of plane optical interconnect techniques have been proposed \cite{Koonath:07}, but these are not compatible with emerging CMOS silicon photonics standards which permit only a single silicon optical wave guiding layer \cite{Bogaerts:05}. In-plane crossing structures, which are CMOS compatible, have been demonstrated \cite{Bogaerts:07}, but these can introduce losses and may be sensitive to fabrication variations.  

Long range communications between waveguides through unguided radiation in the silicon slab have been proposed as an alternate interconnect solution \cite{Naser:OE}. This particular approach uses thin, shallow ridge silicon on insulator waveguides which, when operated in the TM mode, can radiate into the TE modes of the slab  \cite{Webster:07, Kakihara}.  This TE radiation is traditionally considered a loss mechanism, however as it is a coherent process, with appropriate control over the radiation it could be utilised as a resource \cite{Dai:10}.  We have previously shown that it is possible to control the radiation direction \cite{Naser:OE} and also generate directed, collimated beams \cite{Naser:PTL}. The nature of this radiation is quite sensitive to the waveguide geometry and thus may not be robust to fabrication variations. Further, if this radiation is to be used as an interconnect, then the unbound nature of this radiation may lead to undesired interaction with intermediate functional blocks. 

Adiabatic techniques are well known in photonics, principally being invoked when properties of a single waveguide or two waveguide system are changed slowly, for example with adiabatic tapers \cite{Fijol:03}.  Slow changes can also be used to effect population transfer between waveguides through a technique called Coherent Tunnelling Adiabatic Passage (CTAP) which is a spatial analogue of the well-known STIRAP (STImulated Raman Adiabatic Passage) protocol in quantum optics \cite{review}.   CTAP was originally proposed for massive particles in tight-binding systems \cite{Eckert:04, Greentree:04} and then extended to waveguides \cite{Pasp, Longhi:07b}.  CTAP has the advantage that the transport is extremely robust against fluctuations in the coupling between sites. CTAP also has the surprising feature that the population in the intervening site is greatly suppressed, and in the adiabatic, tight-binding limit, is identically zero. This unusual behaviour raises the question of whether CTAP may be exploited to achieve robust long range coupling between waveguides via unbound radiation, but without exciting this radiation.

Here we propose and numerically demonstrate the combination of CTAP and lateral leakage to achieve a new type of coupler. Light guided within one waveguide can be transferred over a long distance to another waveguide through use of an unbound lateral leakage state which is coupled to both waveguides.  Due to the nature of CTAP, this coupling is extremely robust, being relatively independent of coupling length and remarkably, the intermediate radiation is not populated during the coupling. We also show that this technique can be used to bypass an intermediate waveguide without cross-talk. 

This paper is organised as follows:  Section \ref{Section:CTAP} presents a brief over view of the CTAP protocol in the context of optical modes and Section \ref{Section:lateral leakage} reviews lateral leakage and shows how coupling between bound waveguides modes and lateral leakage radiation can be controlled.  Section \ref{Section:2wvg} numerically simulates the CTAP coupling between waveguides and tests the robustness of this technique with varying device length.  Section \ref{Section:3wvg} then shows the bypass of an intermediate waveguide.  Finally, Section \ref{Section:conclusions} discusses the limitations of this specific demonstration and outlines the opportunities for future research on this approach.

\section{Coherent Tunnelling Adiabatic Passage} \label{Section:CTAP}
Coherent Tunnelling Adiabatic Passage (CTAP) is a protocol for transferring population between defined states.  In particular, the transport should be spatial.  It is usual that the modes be in some sense equivalent or discrete, however such restrictions are not always necessary.  

\begin{figure}[t]
\centering
\includegraphics[width=0.85\textwidth]{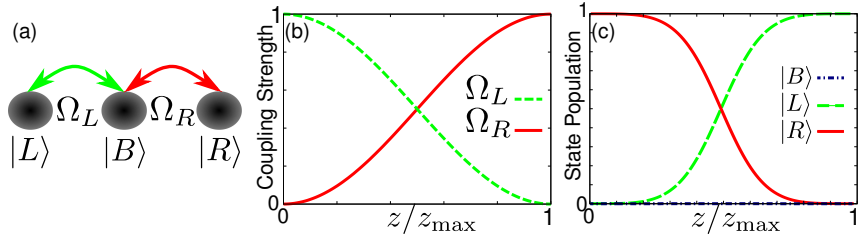}
\caption{The CTAP protocol taking population from $\ket{R}$ to $\ket{L}$:  
(a) A 3 state scheme with two isolated states coupled to a central bus,
(b) counter-intuitive evolution of coupling strengths $\Omega_L$ and $\Omega_R$,
(c) population evolution in states $\ket{L}$, $\ket{R}$ and $\ket{B}$.}
\label{fig:3ctap}
\end{figure}

To illustrate CTAP, consider a three-state system as shown in Fig.~\ref{fig:3ctap}(a). The states $\ket{L}$ and $\ket{R}$ are mutually isolated and can only couple to the common state $\ket{B}$, which acts as a bus. The strength of the couplings between each state and the bus are $\Omega_{L}$ and $\Omega_{R}$. 

The Hamiltonian describing this problem is
\begin{align}
H(z) = \sum_{i = \{L,B,R\}}\beta_i \ket{i}\bra{i} + \Omega_L \ket{B}\bra{L} + \Omega_R \ket{B}\bra{R} + h.c., \label{eq:Ham}
\end{align}

where $\beta_i=k_0n_i$ is the propagation constant for mode $i$ with effective index $n_i$, and $k_0$ is the propagation constant of the free space.

The CTAP protocol is achieved when the couplings are varied in the so called counter-intuitive sequence. This requires $\Omega_L(0) \gg \Omega_R(0)$, and gradual variation in each with increasing $z$ until $\Omega_L(z_{\max}) \gg \Omega_R(z_{\max})$.  There is considerable flexibility in the actual sequence implemented, and popular choices include Gaussian \cite{GRB+1988} and sinusoidal \cite{CH1990} variations, although discontinuities in the controls can also be tolerated under certain conditions \cite{SMM+2007,VG2013}. Here we choose a squared sinusoidal function (Fig.~\ref{fig:3ctap}(b)). The counter-intuitive sequence works by maintaining the system in the null state, which is the supermode (in the limit that all of the $\beta_i$ are equal): 
\begin{align}
\ket{D_0} = \frac{\Omega_R \ket{L} - \Omega_L \ket{R}}{\sqrt{\Omega_L^2 + \Omega_R^2}}. \label{eq:NullState}
\end{align}

Note that this has the desired properties for adiabatic passage, namely that when $\Omega_L \gg \Omega_R$, $\ket{D_0} = \ket{R}$, and when $\Omega_R \gg \Omega_L$, $\ket{D_0} = \ket{L}$.  Provided adiabaticity is preserved, the population in $\ket{B}$ will be identically zero, although the population in $\ket{B}$ only approaches zero when finite mode size is taken into account \cite{CGH+2008}. Here, adiabaticity is defined with respect to the separation (in terms of energy) between $\ket{D_0}$ and the nearest supermode.  Hence the scheme is largely immune to small errors in realisation.  It is also important to recognise that the system is highly insensitive to loss or decoherence mechanisms that act on the bus state, due to the suppressed population there \cite{KCT2008,RK2011,CKR+2012}. 

\section{Thin shallow ridge waveguides and control of lateral leakage} \label{Section:lateral leakage}
Having introduced CTAP in Section \ref{Section:CTAP}, this section introduces thin shallow ridge waveguides and lateral leakage behaviour and shows how this leakage can be controlled for the purpose of implementing a CTAP coupler with this system.

\begin{figure}[t]
\centering
\includegraphics[width=\textwidth]{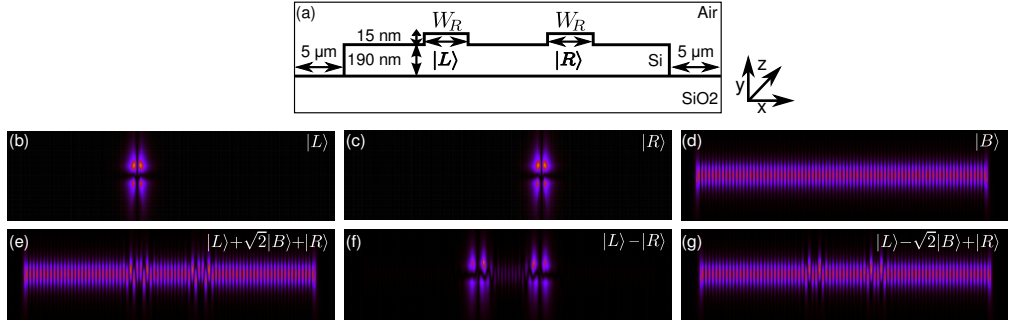}
\caption{
(a) Cross-section of two thin shallow ridges on a silicon slab; simulated modes of the uncoupled structure ($W_{R}=$ 0.7 $\mu$m): (b) and (c) the TM modes $\ket{L}$ and $\ket{R}$ respectively, and (d) the TE slab `bus' mode $\ket{B}$; simulated supermodes of the coupled structure ($W_{R}=$ 1.22 $\mu$m): (e) fundamental ($\ket{B}$ populated), (f) first order ($\ket{B}$ unpopulated), and (g) second order ($\ket{B}$ populated).}
\label{fig:latleak}
\end{figure}

\subsection{Lateral leakage from thin shallow ridge waveguides}
Thin shallow ridge waveguide can be realised using standard CMOS processing and offer highly evanescent modes with low propagation losses. The TM mode can exhibit leakage of power into the laterally radiating TE slab mode. However, this leakage occurs only at the waveguide side walls and at so-called `magic' waveguide widths, the radiation from each side wall cancels \cite{Webster:07}. 

Consider the thin shallow ridge structure of Fig.~\ref{fig:latleak}(a).  To examine the isolated modes of this structure, the waveguide widths were set to the magic width of 0.7 $\mu$m. The guided modes of the system were then simulated using a mode matching method \cite{Thach}. Three simulated guided modes of the system are presented in Fig.~\ref{fig:latleak}(b)-(d). Each had the same effective index and were thus degenerate. Fig.~\ref{fig:latleak}(b) and (c) present the isolated TM modes of the waveguides $\ket{L}$ and $\ket{R}$ respectively.  Fig.~\ref{fig:latleak}(d) presents the TE slab `bus' mode $\ket{B}$.  For simplicity, the slab has been terminated, as illustrated in Fig.~\ref{fig:latleak}(a), and thus the TE slab radiation is in fact a discrete mode with an oscillating standing wave pattern. 

To illustrate the impact of coupling, the widths ($W_{R}$) of waveguides $\ket{L}$ and $\ket{R}$ were set to 1.22 $\mu$m, such that they were strongly and equally coupled to the TE slab $\ket{B}$. The modes of the system were again simulated using mode matching 
and the resulting supermodes, corresponding to the eigenstates of Eq. (\ref{eq:Ham}), are presented in Figs. \ref{fig:latleak}(e)-(g).
Fig.~\ref{fig:latleak}(f) is a supermode with equal population in each of the TM modes $\ket{L}$ and $\ket{R}$ and no population in $\ket{B}$, i.e. the null state, $| L \rangle - | R \rangle$. Whilst Figs. \ref{fig:latleak}(e) and (g) are the supermodes $\ket{L} \pm \sqrt{2}\ket{B} + \ket{R}$ with strong population in the TE slab $\ket{B}$. The three modes of Fig.~\ref{fig:latleak}(e)-(g) are no longer degenerate as the coupling has caused significant splitting of the effective indices of the three modes. 

As discussed in Section \ref{Section:CTAP},  CTAP requires adiabatic transformation of the coupling to transfer the population from $\ket{R}$ of Fig.~\ref{fig:latleak}(c) at the start, into the coupled supermode of Fig.~\ref{fig:latleak}(f) in the middle, and then into $\ket{L}$ of Fig.~\ref{fig:latleak}(b) at the end.  One might consider simply tapering the width of the waveguides to control the coupling, as demonstrated in \cite{Naser:PTL}, however, the modal effective index is sensitive to the waveguide width. For optimal CTAP, it is important that the effective indexes of $\ket{L}$ and $\ket{R}$ remain equal.  Hence an alternative coupling approach is required. 

\begin{figure}[t]
\centering
\includegraphics[width=\textwidth]{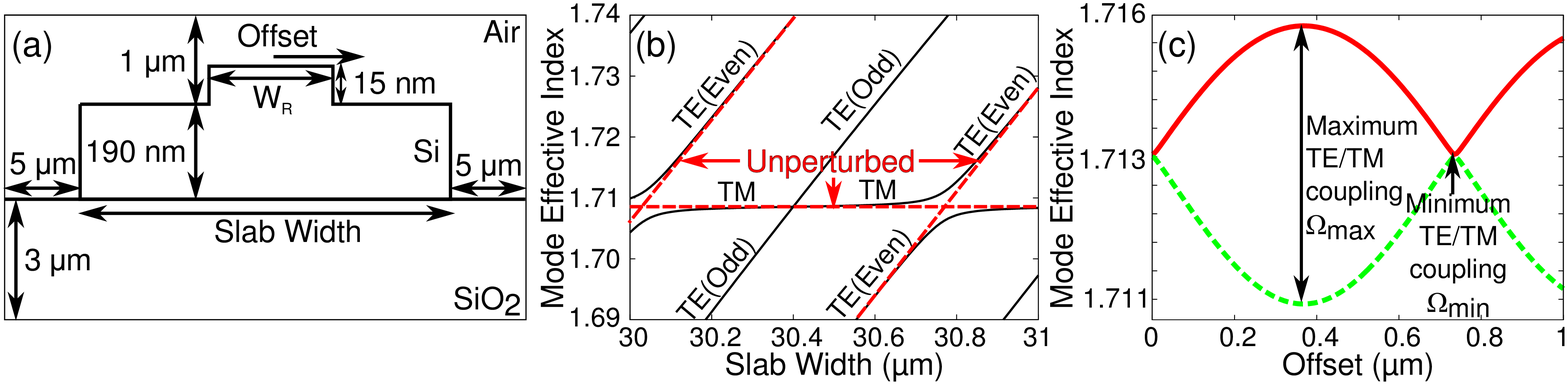}
\caption{
(a) Cross-section of single thin shallow ridge on a silicon slab, 
(b) effective index as a function of slab width for several modes of this system ($W_{R}$ = 1.22 $\mu$m),
(c) effective index as a function of waveguide offset for the phase matched TE and TM modes (slab width = 30.4~$\mu$m).}
\label{fig:1slab}
\end{figure}

\subsection{Control of lateral leakage using waveguide location} \label{Section:control}

An approach to controlling the coupling between the TM waveguides and TE slab that will maintain equal effective indexes for the modes $\ket{L}$ and $\ket{R}$ is suggested by the standing wave pattern of the TE mode as illustrated in Fig.~\ref{fig:latleak}(d). It might be expected that the coupling between the guided TM mode and the TE slab should depend strongly on the lateral location of the thin shallow ridge waveguide.  To establish the effect of waveguide location on coupling between the TM mode and TE slab, the structure of Fig.~\ref{fig:1slab}(a) was modelled.  A single thin shallow ridge waveguide was located on a broad slab.  The width of the thin shallow rib was set to 1.22 $\mu$m such that the TM guided mode should, in principle be strongly coupled to the TE radiation.  

For the particular CTAP protocol we aim to implement it is necessary that $\ket{L}$, $\ket{R}$ and $\ket{B}$ all have the same effective index and are hence degenerate when uncoupled.  Referring to Fig.~\ref{fig:1slab}(a), the waveguide was placed in the centre of the slab, and the slab width was adjusted to find a configuration where the the TM mode and TE slab are degenerate and uncoupled. Mode matching was used to simulate the effective index of the TE and TM modes of this structure as a function of slab width. The results are presented in Fig.~\ref{fig:1slab}(b).  The TE slab mode effective indexes vary with slab width while the index of the TM guided mode remains almost constant. When the TM and TE modes are degenerate, if the symmetry is not matched, the indexes simply cross; however, if the symmetry matches, mode splitting occurs leading to an anti-crossing.  Fig.~\ref{fig:1slab}(b) shows that it is possible to select a slab width where there is a TM guided mode and TE slab mode that are degenerate, but uncoupled at a slab width of 30.4 $\mu$m.

Next the impact of waveguide location on coupling between the TE and TM modes was investigated.  The location of the thin shallow ridge of Fig.~\ref{fig:1slab}(a) was translated laterally across the slab and mode matching was used to simulate the effective indexes of the two supermodes of the system as a function of waveguide offset.  The results are presented in Fig.~\ref{fig:1slab}(c). At 0 nm displacement, the modes are degenerate and uncoupled. As the waveguide was translated, the indexes split, indicating coupling, reaching a maximum at a displacement of 370 nm.  Further displacement decreased the mode splitting until degeneracy was again reached at 740 nm corresponding to a half cycle of the standing wave pattern. These results show that it is indeed possible to control the coupling between the TM and TE modes using waveguide location and this technique could be utilised to implement CTAP with these waveguides. 

\section{Demonstration of long range coupling using CTAP and lateral leakage} \label{Section:2wvg}

Section \ref{Section:control} established that it is possible to control the coupling between localised waveguide modes and distributed slab modes by adjusting their locations. We now show how this coupling control technique can be used to implement a CTAP protocol with thin shallow ridge waveguides.  Specifically, it is shown that power can be adiabatically transferred between two isolated waveguides using TE slab mode radiation as an intermediate bus, but without ever populating this bus.    This section will also test the robustness of this approach by exploring the impact of adjusting device length on the propagation.

\begin{figure}[t]
\centering
\includegraphics[width=0.88\textwidth]{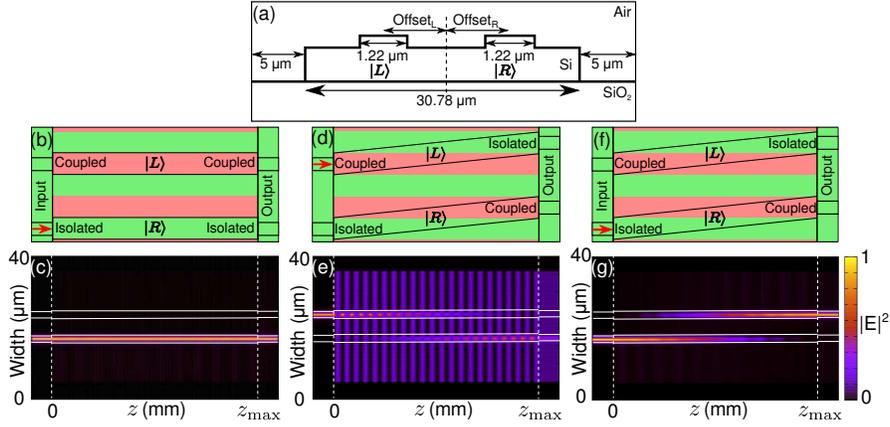}
\caption{ 
(a) Cross-section of two waveguides on a silicon slab; 
(b) plan view of longitudinally invariant \emph{uncoupled} waveguides, excitation on $\ket{R}$; 
(c) optical propagation for \emph{uncoupled} configuration; 
(d) plan view of translated waveguides in \emph{intuitive} CTAP configuration, excitation on $\ket{L}$;   
(e) optical propagation for \emph{intuitive} configuration; 
(f) plan view of translated waveguides in \emph{counter-intuitive} CTAP configuration;
(g) optical propagation for \emph{counter-intuitive} configuration.  
In each case $z_{\max}$ = 10mm.
\label{fig:2prop}
}
\end{figure}

\subsection{CTAP using Lateral Leakage}
Optical propagation in the longitudinally varying structures of this section were simulated using eigenmode expansion (EME) \cite{Naser:OE}. This model rigorously treats the waveguide translation and its impact on the underlying supermodes of the system accounting for radiation.

Fig.~\ref{fig:2prop}(a) presents the cross-section of the geometry under consideration. Two thin shallow ridge waveguides supporting TM modes $\ket{L}$ and $\ket{R}$ were placed on a silicon slab supporting a distributed TE slab mode $\ket{B}$.  The waveguide widths ($W_{R}$) were 1.22 $\mu$m such that the TM and TE modes should be coupled as shown in Section \ref{Section:control}. The slab width was set to 30.78$\mu$m to ensure $\ket{L}$, $\ket{R}$ and $\ket{B}$ had the same effective index. The location of the two waveguides were adjusted to control the coupling between the modes.  Light was coupled into and out of the system through short sections of non-radiating magic width waveguide of width 0.70 $\mu$m.

The configuration of Fig.~\ref{fig:2prop}(b) was considered first with $\ket{L}$ located at 3.47 $\mu$m to be strongly coupled to $\ket{B}$; and $\ket{R}$ located at 3.30 $\mu$m such that it is isolated from $\ket{B}$ throughout propagation.  The separation of 5.38 $\mu$m was expected to be sufficient to ensure no appreciable evanescent coupling directly between $\ket{L}$ and $\ket{R}$. A simulation was performed with $\ket{R}$ excited as indicated by the red arrow on Fig.~\ref{fig:2prop}(b).  The simulation results are presented in Fig \ref{fig:2prop}(c) showing minimal radiation loss from the input region to the propagation region, and no evidence of coupling to either the TE slab $\ket{B}$ or the other TM mode $\ket{L}$.

The structure of Fig.~\ref{fig:2prop}(d) should achieve \emph{intuitive} CTAP coupling.  At the input $\ket{L}$ was located at 3.47 $\mu$m (coupled to $\ket{B}$)  and $\ket{R}$ was at 3.30 $\mu$m (isolated from $\ket{B}$).  However, during propagation, the locations of $\ket{L}$ and $\ket{R}$ were linearly translated, such that at the output, $\ket{L}$ was offset by 3.30 $\mu$m, (isolated from $\ket{B}$), and $\ket{R}$ was at 3.47 $\mu$m (coupled to $\ket{B}$). From Section \ref{Section:control}, linear translation corresponds to sinusoidal evolution of the coupling strength.   A simulation was performed with $\ket{L}$ excited as indicated by the red arrow on Fig.~\ref{fig:2prop}(d).  The results are presented in Fig \ref{fig:2prop}(e).  At the input, light rapidly couples back and forth between $\ket{L}$ and $\ket{B}$.  Mid-way, there is equal and in-phase excitation in both $\ket{L}$ and $\ket{R}$ and rapid coupling to $\ket{B}$ continues with the same coupling length.  At the output, the excitation has transferred to $\ket{R}$ with rapid coupling to $\ket{B}$ still evident.  The output power is split between $\ket{R}$ and $\ket{B}$. This split will be highly sensitive to device length and has been seen in such systems before \cite{Greentree:04,RML+2010}.

The structure of Fig.~\ref{fig:2prop}(f) should achieve \emph{counter-intuitive} CTAP coupling.  This is identical to Fig.~\ref{fig:2prop}(d), however excitation is on $\ket{R}$ as indicated by the red arrow in \ref{fig:2prop}(f). The results are presented in \ref{fig:2prop}(g) which shows smooth transition of the optical power from $\ket{R}$ to $\ket{L}$ without appreciable excitation of $\ket{B}$.  Some slight oscillation is evident, however the rapid, oscillatory coupling to $\ket{B}$ seen in \ref{fig:2prop}(e) are not present. The absence of these oscillations is a major distinguishing feature between CTAP and devices such as directional couplers.

\subsection{Suppression of bus mode excitation and adiabaticity of long range coupling}
\begin{figure}[t]
\centering
\includegraphics[width=0.75\textwidth]{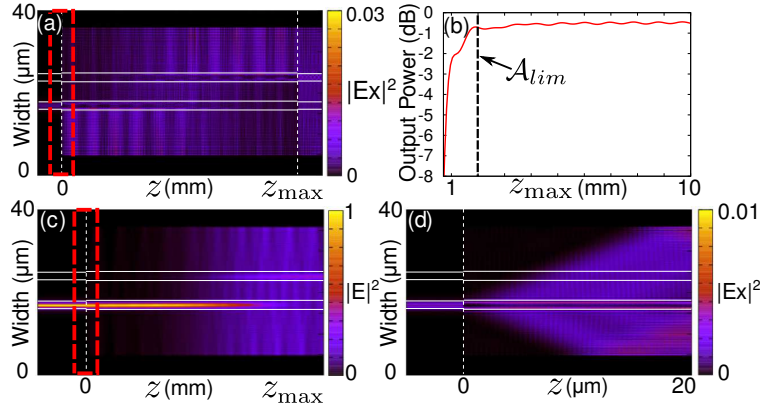}
\caption{
(a) Lateral ($E_{x}$) optical propagation of device with $z_{\max}$ = 10 mm (above adiabatic limit: $A_{lim}$);
(b) $\ket{L}$ output power as a function of device length;
(c) Optical propagation of device with $z_{\max}$ = 500 $\mu$m (below $A_{lim}$);
(d) TE polarised field ($E_{x}$) close to the input termination (common to all simulations).}
\label{fig:2adiab}
\end{figure}

To more closely examine the excitation of $\ket{B}$ during the adiabatic transfer from $\ket{R}$ to $\ket{L}$, the $E_x$ component of the results of Fig.~\ref{fig:2prop}(g) were replotted corresponding to the TE polarisation.  These results are presented in Fig.~\ref{fig:2adiab}(a).  It is evident that there is, in fact, some slight excitation of $\ket{B}$.  There are several effects that can contribute to this residual excitation, including the staircase approximation \cite{VG2013}, finite spatial extent of the modes \cite{CGH+2008}, residual non-adiabaticities in the evolution \cite{Greentree:04}, imperfect initialisation in the null state \cite{JNG+2009} and imperfect coupling of power to the modes of the system at the input and output of the structure.  Fig.~\ref{fig:2adiab}(a) suggests that imperfect coupling to $\ket{R}$ at the input is the dominant source of the population in the TE mode $\ket{B}$, but this effect is deemed negligible for the current demonstration.

The robustness of CTAP protocol was explored by monitoring the coupled power while varying the total device length.  Once in the adiabatic regime, the transport was expected to be largely independent of the exact device length, asymptotically approaching perfect transport.  This contrasts non-adiabatic couplers where the final power would depend critically and periodically on the device length relative to the coupling length.  The structure of Fig.~\ref{fig:2prop}(f) was simulated, but with device length varied from $z_{\max}$ = 0.5 to 10 mm in steps of 50 $\mu$m. Fig.~\ref{fig:2adiab}(b) presents the power coupled from $\ket{L}$ at the output as a function of device length.  For lengths of 2 to 10 mm, the output remains relatively constant indicating adiabatic behaviour while we are operating above the adiabatic limit ($z_{\max} > A_{lim}$) where for this particular structure, $A_{lim}$ is around 1 mm as indicated in Fig.~\ref{fig:2adiab}(b).  The transmission is slightly less than unity and there is a slight ripple evident in the transmission as the length is varied which could be due to the imperfect coupling mentioned above. When the length drops below $A_{lim}$, the transmission begins to drop, falling off dramatically for lengths below 1 mm.  This drop off is due to the device being too short to exhibit adiabatic passage.

Fig.~\ref{fig:2adiab}(c) presents the propagation for the structure of Fig.~\ref{fig:2prop}(f) with $z_{\max}= 500$ $\mu$m. Light input to $\ket{R}$ initially remains isolated from $\ket{B}$, but unlike the behaviour of \ref{fig:2prop}(g), mid-way the light remains in $\ket{R}$ and simply radiates into $\ket{B}$ with minimal coupling to $\ket{L}$. Fig.~\ref{fig:2adiab}(d) presents a highly magnified view of $E_x$ close to the input showing energy naturally radiating from $\ket{R}$ into $\ket{B}$. This coupling occurs at the input where $\ket{R}$ should be isolated, providing evidence that the excitation and isolation of $\ket{R}$ is not perfect.  

\section{CTAP using Lateral Leakage to bypass an intermediate waveguide} \label{Section:3wvg}

\begin{figure}[t]
\centering
\includegraphics[width=0.88\textwidth]{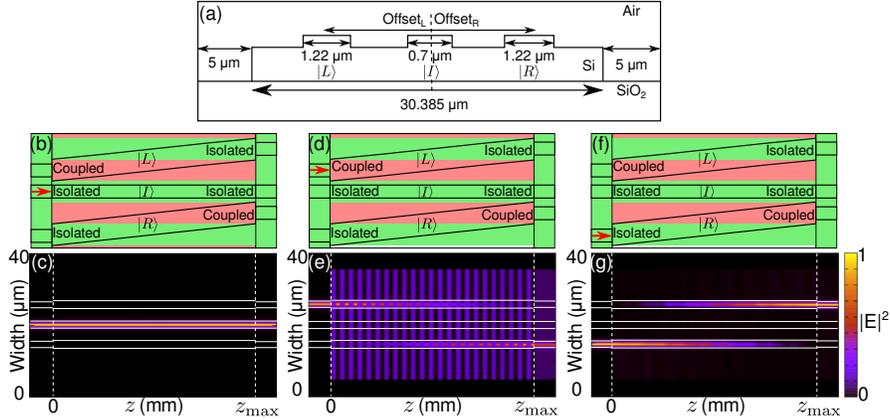}
\caption{
(a) Cross-section of three waveguides on a silicon slab;
(b) plan illustration of bypass CTAP coupler in \emph{uncoupled} configuration with excitation on $\ket{I}$;
(c) optical propagation for \emph{uncoupled} configuration;
(d) plan illustration of bypass CTAP coupler in \emph{intuitive} configuration with excitation on $\ket{L}$;
(e) optical propagation for \emph{intuitive} configuration;
(f) plan illustration of bypass CTAP coupler in \emph{counter-intuitive} configuration with excitation on $\ket{R}$;
(g) optical propagation for \emph{counter-intuitive} configuration.
In each case $z_{\max}$ = 10mm.
\label{fig:3geo}
}
\end{figure}

Whilst the demonstration of Section \ref{Section:2wvg} is interesting, this does not provide the functionality for long range interconnections across a complex planar system. We now show this functionality by demonstrating that CTAP using lateral leakage can bypass an intermediate waveguide. 

The structure of Fig.~\ref{fig:3geo}(a) is similar to that of Fig.~\ref{fig:2prop}(a), but has an additional intermediate waveguide, $\ket{I}$, inserted at the centre. The waveguide supporting $\ket{I}$ was maintained at the magic width throughout propagation in order to isolate it from the slab mode, $\ket{B}$ irrespective of its location.  The width of the slab was altered to 30.385 $\mu$m to ensure that $\ket{B}$ was phase matched to $\ket{L}$ and $\ket{R}$. The offset on $\ket{L}$ and $\ket{R}$ were $\pm$ 5.652 $\mu$m to achieve isolation and $\pm$ 5.484 $\mu$m to achieve coupling to $\ket{B}$. This increased offset from the centre aimed to ensure no evanescent coupling between $\ket{L}$, $\ket{R}$ and $\ket{I}$. The structure was configured as in Fig.~\ref{fig:3geo}(b) such that at the input, $\ket{R}$ was isolated and $\ket{L}$ was coupled to $\ket{B}$ and followed the same counter-intuitive translation as in Fig.~\ref{fig:2prop}(f). The intermediate state $\ket{I}$ was not translated, however it would be expected that translation of the intermediate waveguide would not impact the performance of the device. Each of the three waveguides was interfaced at the input and output to non-radiating magic width waveguides. The structure was simulated as described in Section \ref{Section:2wvg}.

The first simulation tested the isolation of $\ket{I}$ from $\ket{L}$, $\ket{R}$ and $\ket{B}$. Optical power was input to the intermediate waveguide as indicated by the red arrow in Fig.~\ref{fig:3geo}(b).  Fig.~\ref{fig:3geo}(c) presents the simulated results showing that light remains confined to the intermediate waveguide without any evidence of coupling.  Next the \emph{intuitive} coupling case of Fig.~\ref{fig:3geo}(d) was simulated and is presented in Fig.~\ref{fig:3geo}(e). These results can be compared to Fig.~\ref{fig:2prop}(e) exhibiting similar population oscillations. Importantly, there is no evident coupling into $\ket{I}$, as expected since it is at the magic width and should be isolated from the TE slab. 

Finally, the \emph{counter-intuitive} coupling case of Fig.~\ref{fig:3geo}(f) was simulated and the results are presented in Fig.~\ref{fig:3geo}(g). Comparing these results to Fig.~\ref{fig:2prop}(g), it can be seen that again adiabatic passage without appreciable population in either the bus mode, $\ket{B}$, or intermediate waveguide, $\ket{I}$, has been achieved. Slight pulsing of the light is again observed in Fig.~\ref{fig:3geo}(g) similar to that of Fig.~\ref{fig:2prop}(g).  These simulations confirm that this adiabatic coupling structure is indeed capable of transferring an optical signal from one waveguide to another, bypassing an intermediate waveguide using the TE slab mode as a type of bus, but without ever populating this bus. 

\section{Conclusions} \label{Section:conclusions}

We have described a new concept for adiabatic transfer of power between two thin shallow ridge waveguides and proved this concept using rigorous numerical simulation.  The power transfer occurs by coupling each waveguide to a laterally distributed slab mode which acts as an optical bus. The novelty of our demonstrated concept is that due to the nature of the Coherent Tunnelling Adiabatic Passage (CTAP) protocol employed, power is robustly transferred from one waveguide to the other without ever populating the intermediate optical bus. The distributed nature of the bus allows the coupling to be long-range, exceeding evanescent interaction distances and indeed extending beyond nearest neighbour interactions. We have demonstrated this feature by showing that our CTAP coupler can bypass an intermediate waveguide. Since the bus state population is minimal, and the CTAP protocol is highly robust we would expect the transfer to be insensitive to other intervening structures or imperfections of the slab. 

We propose that this new coupling technique could have a potential application as an interconnect mechanism across complex integrated optical systems. However, before this approach can be taken beyond the proof of concept stage, there are limitations and possible extensions that should be explored.  

A significant restriction of our demonstration is that in our demonstration the transport was via a discrete mode of the slab.  This has obvious limitations as it imposes a restriction on the properties of the whole slab, rather than just the slab in the vicinity of the active waveguides.  However, there are STIRAP/CTAP protocols that operate using multiple intermediate states \cite{VSB1998, DSH+2009} and even via a continuum \cite{PYH2005}, again with minimal occupation of those intermediate states.  Since our CTAP protocol is a direct analogy of STIRAP we are confident that similar approaches could be employed to eliminate the dependence on the properties of the discrete modes of the slab. 

In our demonstration of adiabatic transfer bypassing an intermediate waveguide only two waveguides were coupled to the slab at any one time with the third intermediate waveguide maintained at the magic width and hence uncoupled from the slab at all times.  It would be of interest to explore cases where more than two waveguides are coupled to the bus simultaneously, for example topologies equivalent to the tripod and multi-pod schemes from STIRAP. These schemes have been proposed for applications such as geometric gates \cite{USB1999} and multiple-recipient adiabatic passage \cite{GDH2006}, which cannot be realised without some form of non-nearest neighbour coupling, such as has been outlined here.

\section*{Acknowledgments}
A.P.H, T.G.N and A.M. acknowledge the support of the Australian Research Council (ARC) Centre of Excellence Funding (CE110001018), A.P.H. acknowledges Robert and Josephine Shanks Scholarship, T.G.N. acknowledges support from the ARC APD fellowship (DP1096153), A.D.G. acknowledges the ARC for financial support (Grants No. DP0880466 and No. DP130104381).


\begin{thebibliography}{99}

\bibitem{Bogaerts:05}
W.~Bogaerts, R.~Baets, P.~Dumon, V.~Wiaux, S.~Beckx,
D.~Taillaert, B.~Luyssaert, J.~Van Campenhout, P.~Bienstman, and D.~Van Thourhout,
\enquote{Nanophotonic Waveguides in Silicon-on Insulator Fabricated With CMOS Technology,} J. Lightwave Technol. \textbf{23}, 401, (2005)

\bibitem{Ding:12}
R.~Ding, T.~Baehr-Jones, T.~Pinguet, J.~Li, N.~Harris, M.~Streshinsky, L.~He, A.~Novack, E.~Lim, T.~Liow, H.~Teo, G.~Lo, and M.~Hochberg, 
\enquote{A Silicon Platform for High-Speed Photonics Systems,} in Optical Fiber Comm. Conf., (2012)

\bibitem{Koos:09}
C.~Koos, P.~Vorreau, T.~Vallaitis, P.~Dumon, W.~Bogaerts, R.~Baets, B.~Esembeson, I.~Biaggio, T.~Michinobu, F.~Diederich, W.~Freude and J.~Leuthold,
\enquote{All-optical high-speed signal processing with silicon-organic hybrid slot waveguides,}
Nat. Photonics \textbf{3}, 216--219, (2009)

\bibitem{Politi:08}
A.~Politi, M.~Cryan, J.~Rarity, S.~Yu, J.~O'Brien,
\enquote{Silica-on-Silicon Waveguide Quantum Circuits,}
Science \textbf{320}, 646--649, (2008)

\bibitem{Koonath:07}
P.~Koonath and B.~Jalali, 
\enquote{Multilayer 3-D photonics in silicon,}
Opt. Express \textbf{15}, 12686--12691, (2007)

\bibitem{Bogaerts:07}
W.~Bogaerts, P.~Dumon, D.~Thourhout, and R.~Baets, 
\enquote{Low-loss, low-cross-talk crossings for silicon-on-insulator nanophotonic waveguides,} Opt. Lett.  \textbf{32}, 2801--2803, (2007)

\bibitem{Naser:OE}
N.~Dalvand, T.~G.~Nguyen, R.~S.~Tummidi, T.~L.~Koch, A.~Mitchell, 
\enquote{Thin-ridge Silicon-on-Insulator waveguides with directional control of lateral leakage radiation,} Opt. Express \textbf{19}, 5635--5643, (2011)

\bibitem{Webster:07}
M.~Webster, R.~Pafchek, A.~Mitchell, and T.~Koch, \enquote{Width dependence of inherent TM-mode lateral leakage loss in silicon-on-insulator ridge waveguides,} IEEE Photon. Technol. Lett. \textbf{19}(6), 429--431, (2007)

\bibitem{Kakihara}
K.~Kakihara, K.~Saitoh, M.~Koshiba, 
\enquote{Generalized Simple Theory for Estimating Lateral Leakage Loss Behavior in Silicon-on-Insulator Ridge Waveguides,} 
J. Light. Tech. \textbf{27}(23), 5492--5499, (2009)

\bibitem{Dai:10}
D.~Dai, Z.~Wang, N.~Julian, and J.~Bowers, 
\enquote{Compact broadband polarizer based on shallowly-etched silicon-on-insulator ridge optical waveguides,} 
Opt. Express \textbf{18}, 27404--27415, (2010)

\bibitem{Naser:PTL}
N.~Dalvand, T.~G.~Nguyen, T.~L.~Koch, A.~Mitchell, 
\enquote{Thin Shallow-Ridge Silicon-on-Insulator Waveguide Transitions and Tapers,} Photon. Technol. Lett. \textbf{25}, 163--166, (2013)

\bibitem{Fijol:03}
J.~J.~Fijol, E.~E.~Fike, P.~B.~Keating, D.~Gilbody, J.~J.~LeBlanc, S.~A.~Jacobson, W.~J.~Kessler, M.~B.~Frish,
\enquote{Fabrication of silicon-on-insulator adiabatic tapers for low-loss optical interconnection of photonic devices,} 
Proc. SPIE 4997, Phot. Pack. and Integr. III Conf. \textbf{157}, (2003)

\bibitem{review}
N.V.~Vitanov, T.~Halfmann, B.W.~Shore, K.~Bergmann,
\enquote{Laser-inducted population transfer by adiabatic passage techniques,}
Annu. Rev. Phys. Chem. \textbf{52}, 763209, (2001)

\bibitem{Eckert:04}
K.~Eckert, M.~Lewenstein, R.~Corbal\'{a}n, G.~Birkl, W.~Ertmer, and J.~Mompart,
\enquote{Three-level atom optics via the tunneling interaction,} Phys. Rev. A \textbf{70}, 023606, (2004)

\bibitem{Greentree:04}
A.~D. Greentree, J.~H. Cole, A.~R. Hamilton, and L.~C.~L. Hollenberg,
\enquote{Coherent electronic transfer in quantum dot systems using adiabatic passage,}
Phys. Rev. B \textbf{70}, 235317, (2004)

\bibitem{Pasp}
E.~Paspalakis,
\enquote{Adiabatic three-waveguide directional coupler,}
Opt. Commun. \textbf{258}, 30--34, (2006)

\bibitem{Longhi:07b}
S.~Longhi, G.~Della~Valle, M.~Ornigotti, and P.~Laporta, 
\enquote{Coherent tunneling by adiabatic passage in an optical waveguide system,} 
Phys. Rev. B \textbf{76}, 201101 (2007)

\bibitem{GRB+1988}
U.~Gaubatz, P.~Rudecki, M.~Becker, S.~Schiemann, M.~K\"{u}lz, K.~Bergmann, 
\enquote{Population switching between vibrational levels in molecular beams,}
Chem. Phys. Lett. \textbf{149}, 463--468, (1988)

\bibitem{CH1990}
C. E.~Carroll and F. T.~Hioe, 
\enquote{Adiabatic processes in three-level systems,} 
Phys. Rev. A \textbf{42}, 1522, (1990)

\bibitem{SMM+2007}
E.~A.~Shapiro, V.~Milner, C.~Menzel-Jones, and M.~Shapiro, 
\enquote{Piecewise Adiabatic Passage with a Series of Femtosecond Pulses,} 
Phys. Rev. Lett. \textbf{99}, 033002, (2007)

\bibitem{VG2013}
J.~A.~Vaitkus, and A.~D.~Greentree,
\enquote{Digital three-state adiabatic passage,}
Phys. Rev. A \textbf{87}, 063820, (2013)

\bibitem{CGH+2008}
J.~H.~Cole, A.~D.~Greentree and L.~C.~L.~Hollenberg, and S.~Das Sarma,
\enquote{Spatial adiabatic passage in a realistic triple well structure,}
Phys. Rev. B \textbf{77}, 235418, (2008)

\bibitem{KCT2008}
I.~Kamleitner, J.~Cresser, and J.~Twamley, 
\enquote{Adiabatic information transport in the presence of decoherence,}
Phys. Rev. A \textbf{77}, 032331, (2008)

\bibitem{RK2011}
J.~Rech and S.~Kehrein, 
\enquote{Effect of Measurement Backaction on Adiabatic Coherent Electron Transport,}
Phys. Rev. Lett. \textbf{106}, 136808, (2011)

\bibitem{CKR+2012}
K.~Chung, T. J.~Karle, M.~Rab, A. D.~Greentree, and S.~Tomljenovic-Hanic, 
\enquote{Broadband and robust optical waveguide devices using coherent tunnelling adiabatic passage,}
Opt. Express \textbf{20}, 23108, (2012)

\bibitem{Thach}
T.~Nguyen, R.~Tummidi, T.~Koch, and A.~Mitchell, 
\enquote{Rigorous modeling of lateral leakage loss in SOI thin-ridge waveguides and couplers,}
Photon. Technol. Lett., IEEE \textbf{21}, 486--488 (2009)

\bibitem{Ravi}
R.~Tummidi, T.~Nguyen, A.~Mitchell, and T.~Koch, \enquote{An ultra-compact waveguide polarizer based on `anti-magic' widths,} 
in \enquote{Group IV Photonics (GFP), 8th IEEE Int. Conf.,} 104--106, (2011)
  
\bibitem{RML+2010}
R.~Rahman, R.~P.~Muller, J.~E.~Levy, M.~S.~Carroll, G.~Klimeck, A.~D.~Greentree, and L.~C.~L.~Hollenberg, 
\enquote{Coherent electron transport by adiabatic passage in an imperfect donor chain,}
Phys. Rev. B \textbf{82}, 155315, (2010)

\bibitem{Jalali:06}
B.~Jalali, and S.~Fathpour,
\enquote{Silicon Photonics,}
J. Lightwave Technol. \textbf{24}, 4600--4615, (2006)

\bibitem{Baba:13}
T.~Baba, S.~Akiyama, M.~Imai, N.~Hirayama, H.~Takahashi, Y.~Noguchi, T.~Horikawa, and T.~Usuki, 
\enquote{50-Gb/s ring-resonator-based silicon modulator,}
Opt. Express \textbf{21}, 11869--11876, (2013)

\bibitem{Meany:12}
T.~Meany, M.~Delanty, S.~Gross, G.~D.~Marshall, M.~J.~Steel, M.J.~Withford,
\enquote{Non-classical interference in integrated 3D multiports,}
Opt. Express \textbf{20}, 26895--26905, (2012)

\bibitem{Peters:07}
T.~Peters and T.~Halfmann, 
\enquote{Stimulated Raman adiabatic passage via the ionization continuum in helium: Experiment and theory,}
Opt. Commun. \textbf{271}, 475--486, (2007)

\bibitem{VSB1998}
N.~V.~Vitanov, B.~W.~Shore, and K.~Bergmann, 
\enquote{Adiabatic population transfer in multistate chains via dressed intermediate states,}
Eur. Phys. J. D \textbf{4}, 15, (1998)

\bibitem{DSH+2009}
F. Dreisow, A. Szameit, M. Heinrich, R. Keil, S. Nolte, A. T\"{u}nnermann, and S. Longhi, 
\enquote{Adiabatic transfer of light via a continuum in optical waveguides,} 
Opt. Lett. \textbf{34}, 2405--2407, (2009)

\bibitem{PYH2005}
T.~Peters, L.~P.~Yatsenko, and T.~Halfmann, 
\enquote{Experimental Demonstration of Selective Coherent Population Transfer via a Continuum,}
Phys. Rev. Lett. \textbf{95}, 103601, (2005)

\bibitem{USB1999}
R. G.~Unanyan, B. W.~Shore, and K.~Bergmann, 
\enquote{Laser-driven population transfer in four-level atoms: Consequences of non-Abelian geometrical adiabatic phase factors,}
Phys. Rev. A \textbf{59}, 2910, (1999)

\bibitem{GDH2006}
A.~D.~Greentree, S.~J.~Devitt, and L.~C.~L.~Hollenberg, 
\enquote{Quantum-information transport to multiple receivers,}
Phys. Rev. A \textbf{73}, 032319, (2006)

\bibitem{JNG+2009}
L.~M.~Jong, A.~D.~Greentree, V.~I.~Conrad, L.~C.~L.~Hollenberg and D.~N.~Jamieson, 
\enquote{Coherent tunneling adiabatic passage with the alternating coupling scheme,} 
Nanotech. \textbf{20}, 405402, (2009)

\end{thebibliography}
\end{document}